\documentstyle[aaspp4]{article}

\def\degrees{$^\circ$}

\def\spose#1{\hbox to 0pt{#1\hss}}
     
\def\={\overline}

\newcount\notenumber
\newcount\eqnumber
\newcount\fignumber
\newbox\abstr
\newbox\figca

\def\etal{{\it et al. }}
\def\eg{{\it e.g., }}
\def\ie{{\it i.e., }}
\def\cf{{\it cf. }}

\def\note#1{\footnote{$^{\the\notenumber}$}{#1}\global\advance\notenumber by 1}
\def\foot#1{\raise3pt\hbox{\eightrm \the\notenumber}
     \hfil\par\vskip3pt\hrule\vskip6pt
     \noindent\raise3pt\hbox{\eightrm \the\notenumber}
     #1\par\vskip6pt\hrule\vskip3pt\noindent\global\advance\notenumber by 1}

\def\Dt{\spose{\raise 1.5ex\hbox{\hskip3pt$\mathchar"201$}}}    
\def\dt{\spose{\raise 1.0ex\hbox{\hskip2pt$\mathchar"201$}}}    

\def\new{{\rm\chaphead\the\eqnumber}\global\advance\eqnumber by 1}
\def\ref#1{\advance\eqnumber by -#1 \chaphead\the\eqnumber
     \advance\eqnumber by #1 }
\def\last{\advance\eqnumber by -1 {\rm\chaphead\the\eqnumber}\advance
     \eqnumber by 1}
\def\eqnam#1{\xdef#1{\chaphead\the\eqnumber}}
     
\def\nfig{\chaphead\the\fignumber\global\advance\fignumber by 1}
\def\nfiga#1{\chaphead\the\fignumber{#1}\global\advance\fignumber by 1}
\def\rfig#1{\advance\fignumber by -#1 \chaphead\the\fignumber
     \advance\fignumber by #1}
\def\refindent{\par\noindent\parskip=4pt\hangindent=3pc\hangafter=1 }

\def\apj#1#2#3{\refindent#1,  {ApJ,\ }{#2}, #3}
\def\apjsup#1#2#3{\refindent#1,  {ApJS\ }{#2}, #3}
\def\aasup#1#2#3{\refindent#1,  { AA Sup.,\ }{#2}, #3}

\def\apjlett#1#2#3{\refindent#1,  { ApJL,\  }{#2}, #3}
\def\mn#1#2#3{\refindent#1,  { MNRAS,\ }{#2}, #3}

\def\annrev#1#2#3{\refindent#1, { ARA \& A,\ }
{#2}, #3}
\def\aj#1#2#3{\refindent#1,  { AJ,\  }{#2}, #3}

\def\aa#1#2#3{\refindent#1,  { AA,\ }{#2}, #3}

\def\pasp#1#2#3{\refindent#1,  { PASP,\ }{#2}, #3}

\def\refbook#1{\refindent#1}

\def\ltsim{\mathrel{\spose{\lower 3pt\hbox{$\mathchar"218$}}
     \raise 2.0pt\hbox{$\mathchar"13C$}}}
\def\gtsim{\mathrel{\spose{\lower 3pt\hbox{$\mathchar"218$}}
     \raise 2.0pt\hbox{$\mathchar"13E$}}}
     
\def\apequal{\mathrel{\spose{\lower 1pt\hbox{$\mathchar"218$}}
     \raise 2.0pt\hbox{$\mathchar"218$}}}
\newbox\grsign \setbox\grsign=\hbox{$>$} \newdimen\grdimen \grdimen=\ht\grsign
\newbox\simlessbox \newbox\simgreatbox
\setbox\simgreatbox=\hbox{\raise.5ex\hbox{$>$}\llap
     {\lower.5ex\hbox{$\sim$}}}\ht1=\grdimen\dp1=0pt
\setbox\simlessbox=\hbox{\raise.5ex\hbox{$<$}\llap
     {\lower.5ex\hbox{$\sim$}}}\ht2=\grdimen\dp2=0pt

\def\Kp{K$^\prime$\ }
\def\etal{{\it et al.\ }}

\begin{document}

\def\Kp{K$^\prime$\ }
\def\etal{{\it et al.\ }}
\title{\bf 
Evidence for an Intervening Stellar Population Toward the Large Magellanic
Cloud$^1$}
\author{Dennis Zaritsky and D.N.C. Lin}
\bigskip
\affil{UCO/Lick Observatory and Board of Astronomy and Astrophysics,
Univ. of California at Santa Cruz, Santa Cruz, CA 95064. Email:
dennis@ucolick.org, lin@ucolick.org}
\medskip
$^1$ Lick Bulletin No. 1364
\vskip 1in

\setcounter{footnote}{3}
\begin{abstract}
We identify a vertical extension of the red clump stars in
the color magnitude diagram (CMD) of a section of the Large Magellanic
Cloud (LMC). The
distribution of stars in this extension is indistinguishable 
in the $U$, $B$, $V$, and $I$ bands --- confirming that the detection
is real and placing a strong constraint on models of this stellar
population. After subtracting the principal red clump component,
we find a peak in the residual
stellar distribution that is $\sim$ 0.9 mag brighter than the peak
of the principal red clump distribution. 
We consider and reject the following possible explanations for this 
population: 
inhomogeneous reddening, Galactic disk stars,
random blends of red clump stars, correlated blends of red clump
stars (binaries), evolution of the 
red clump stars, and red clump stars from a
younger LMC stellar population.
Combinations of these effects cannot be ruled out
as the origin of this stellar population.

{\it A natural interpretation of this new population is 
that it consists of red clump stars that are closer to us than those in
the LMC.} We derive a distance for this population of $\sim$ 33 to 35 kpc,
although the measurement is sensitive to the
modeling of the LMC red clump component. We find corroborating evidence
for this interpretation in Holtzman \etal's (1997)
{\it Hubble Space Telescope} CMD
of the LMC field stars. The derived distance
and projected angular surface density of these stars relative to the
LMC stars
($\ltsim$ 5 to 7\%)  are 
consistent with (1) models that
attribute the observed microlensing lensing optical depth (Alcock
\etal 1997) to a distinct foreground stellar population (Zhao 1997)
and (2) tidal models of the interaction between the
LMC and the Milky Way (Lin, Jones, \& Klemola 1995).
The first result suggests
that the Galactic halo may contain few, if 
any, purely halo MACHO objects. 
The second result
suggests that this new population may be evidence of a tidal tail from 
the interaction between the LMC and the Galaxy (although other
interpretations, such as debris from the LMC-SMC interaction, are possible).
We conclude that the standard assumption of a smoothly distributed
halo population 
out to the LMC cannot
be substantiated without at least 
a detailed understanding of several of the following:
red clump stellar
evolution, binary fractions, binary mass ratios, the spatial
correlation of stars within the LMC,
possible variations in the stellar populations of satellite galaxies,
and differential reddening --- all of which
are highly complex. 

\end{abstract}

\section{Introduction}

Identifying structure in the Galactic halo
is critical to understanding the
formation of the halo and to interpreting the microlensing
observations that provide constraints on the composition of Galactic dark
matter. Halo structures, such as tidal streamers, are the relics
of the process of galaxy formation and, as 
envisioned in hierarchical models of the growth
of structure, will testify to past and present accretion 
events. Furthermore, a concentration of stars in the halo will alter the
microlensing optical depth along that line of sight 
and affect subsequent interpretations
regarding the density of halo MACHO objects. 
These structures are difficult to identify because 
their stellar surface density 
is likely to be a small.
Therefore, large photometric surveys are
required to identify these systems against the
backdrop of halo stars (Johnston, Hernquist, \& Bolte 1996).
Our observations of the Large Magellanic Cloud (LMC)
provide $UBVI$ data for nearly 1 million stars (Zaritsky, Harris,
\& Thompson 1997) and an opportunity to search for intervening
populations
along the line of sight to the LMC.

As satellite galaxies or proto-galactic fragments interact with
the Galaxy, tidal debris will be strewn through the halo. Identifying
these remains, particularly if the interaction occurred many Gyr
ago, is difficult, but some evidence for such events exists.
Two particular examples of ongoing interactions between the Galaxy and 
its satellites are well established: the gaseous Magellanic Stream
(Mathewson, Schwarz, \& Murray 1977)
and the tidal extension of the Sagittarius dwarf galaxy
(Ibata, Gilmore, \& Irwin 1995; Mateo \etal 1996).
Somewhat more tentative is the
inference that ancient interactions led to
the set of satellite galaxies and globular clusters
that share two preferred orbital planes around the Galaxy 
(Kunkel 1979; Lynden-Bell 1982; Majewski 1994). These orbital planes are
presumably also populated by individual stars that were tidally stripped from
the original parent satellite. Attempts to detect 
stars in the Magellanic Stream (\eg Br\"uck \& Hawkins 1983), 
have not yet yielded 
positive results, leading some (\eg Meurer, Bicknell, \& Gingold 1985; 
Moore \& Davis 1994) to propose models
in which the stream material is removed from the Clouds by ram pressure 
rather than by tidal forces. As the known coherent structures
are studied further, and other structures are discovered, 
we will be able develop a clearer
picture of the physics that led to the formation of the outer Galactic 
halo and the evolution of the Galactic satellites. 

A key development over the past few years in the effort to determine
the nature of Galactic dark matter has been the
systematic discovery of microlensing events toward the LMC (cf. Alcock \etal
1996). However, the interpretation of those results hinges on models
of the stellar distribution along the the line of sight.
Zhao (1997) has demonstrated that the microlensing optical depth 
may be dominated by 
tidal debris from the interaction of the LMC with the Galaxy, rather
than by halo MACHOS, even if such tidal material has a stellar surface
density
of only a few percent of the LMC stellar surface density.
Therefore, a detailed determination of the stellar distribution along the
line of sight to the LMC is essential before reaching conclusions
regarding the origin, density, and mass 
of the lensing population.

In this paper, we present observations that can be interpreted as
evidence for a stellar population at a distance of $\sim$ 35 kpc
toward the LMC.
In \S2 we discuss the data, possible alternative interpretations of
observations, and the line-of-sight
stellar density distribution that we derive from these data. Although
we cannot definitely exclude alternative interpretations, the
intervening population hypothesis is entirely consistent with the
data. 
Other interpretations (\eg that there are a large number of red clump binaries
in the LMC or that there is a problem with current models of red clump stellar evolution)
are also physically interesting.
In \S3 we discuss the potential effect of this population on 
conclusions drawn from the observed microlensing of
LMC stars, 
the possible origins of such a stellar population,
and the implications for dynamical models of the tidal interaction between the
LMC and the Galaxy or between the SMC and LMC.
In particular, the location and projected density of the
proposed foreground 
stellar population are consistent with a model in which they are responsible
for the
microlensing detection rate toward the LMC (as discussed hypothetically
by Zhao 1997) and 
with a model of the tidal 
interaction between the Galaxy and the LMC 
(Lin, Jones, \& Klemola 1995).
Finally, we discuss some future observations that may
discriminate between the various 
interpretations of the current data.

\section{The Observations and Apparent Stellar Line-of-Sight Distribution}

Our data come from digital, $UBVI$, drift scan images
of a 2\degrees\ $\times$ 1.5\degrees\ region located
$\sim$ 2\degrees\ northwest
of the center of the Large Magellanic Cloud. These data are 
part of an ongoing survey conducted at the Las Campanas 1-m
Swope telescope that is designed to image
the central 8\degrees\ by 8\degrees\ of the LMC and
4\degrees\ by 4\degrees\ of the SMC in the $UBVI$ bandpasses using
the GCC drift-scan camera (Zaritsky, Shectman, 
\& Bredthauer 1996). The effective exposure time
is defined by the time required for the sky to drift across
the field-of-view of the stationary telescope ($\sim$
240 sec at the declination of the Clouds). The CCD 
has a 0.7 arcsec pixel$^{-1}$ scale, and the typical seeing at the
telescope is between 1.2 and 1.8 arcsec. These data were obtained
during Nov. 13-23, 1995.
The details of the data reduction, the photometric and astrometric
precision, and the photometric
completeness are discussed by Zaritsky, Harris, \& Thompson (1997).

We apply additional criteria to the existing catalog of
stellar photometry. First, we only analyze stars for which 
magnitudes are available in all four filter bands and for which
the $V$ magnitude is brighter than the 50\%
completeness limit, $m_V =$ 21 mag (see Zaritsky, Harris,
\& Thompson for details). Second,
to minimize the number of stars that may have
been mismatched between images taken in different filter bandpasses,
we fit blackbody curves to all of the remaining stars.
Because blackbody curves does not describe the full complexity
of a stellar spectrum (\eg they lack absorption lines), 
fits with $\chi^2 > 1$ are generally acceptable.
We interactively determine that
stars with fits that have $\chi^2 \ltsim 30$ are acceptable and
remove stars from the sample with best fits that have $\chi^2 > 30$.
The fraction of ``stars'' removed from the sample
due to this criterion is small (0.02). 

To examine the distribution of stars along the line
of sight, 
we use the most distinctive and common stellar population in 
the bright ($V > 21$) portion of the LMC color magnitude diagram (CMD) --- 
the red clump stars --- as a tracer of the full stellar population. 
We assume that red
clump stars are standard candles.
This assumption has been made by other investigators to examine
the structure of the SMC (Hatzidimitriou and Hawkins 1989) and the 
Galactic bar (Stanek \etal 1997). 
These papers, and the references therein, argue that
the luminosity of the red clump is relatively insensitive to age
and metallicity if the stellar population is older than 1 Gyr
(models predict luminosity changes $\ltsim 0.6$ mag for 
a given metallicity; Sweigart 1987).

Even if one accepts that red clump stars are fair
standard candles, this
approach to deriving the line of sight stellar density
has some difficulties. For example,
red clump
stars and stars at the base of the red giant branch can be confused
because the
two populations have similar $B-V$ colors. To increase the distinction
between these populations, 
we define a new color, constructed from a combination of the
available four filter photometry,
that maximally separates stars along
the color axis in the red clump and the red giant branch portion
of the CMD. 
We found the $U-V$,$B-I$ color-color
plot to provide the greatest discrimination between the two
evolutionary
phases. Other choices, such as $U-B$,$V-I$, do not produce as 
significant a color variation as $U-V$,$B-I$ for these stars. To align the
maximum dispersion of the stellar distribution along one axis, 
we apply a rotation to the color-color space. The axis with maximum
dispersion
defines our new color, $C \equiv 0.565(B-I) + 0.825(U-V + 1.15)$.
We do not correct for extinction
and justify this omission below.

The resulting Hess diagrams for the $B$, $V$, and $I$ data near the
red clump region
are presented in Figure 1. The diagrams are created by modeling
each star as a two-dimensional Gaussian with dispersions given by
the photometric errors. All of the stars are summed to generate 
the density-coded CMD (or Hess diagram).
In the $V$-band panel, we label the four features of interest:
(1) the central red clump (RC), (2), the vertical extension of
the red clump (VRC), (3) the red giant branch bump (RGBB; see King \etal 1985;
Fusi Pecci \etal 1990), and (4) the 
giant branch (the unresolved combination of the red giant and 
asymptotic giant branches). Our claim, which we will attempt to 
justify to the reader, is that the VRC is real and that it consists
of foreground red clump stars. We discuss
the relative merits of various interpretations of the VRC next.

\noindent
{\it a) The VRC is spurious.}

Is the VRC truly a distinct component or merely an artifact of the 
photometric uncertainties?
We argue that the VRC is real due to its presence 
in the data from each of the four filter bands.
To quantify this statement, we consider the magnitude distribution of
stars in a vertical (\ie constant color) band of the CMD
that includes the red clump stars. The boundaries of this region are 
taken as $2.85 < C < 3.57$, where
$C$ is the color defined above. This color range is
selected to straddle the centroid of the RC distribution.
The local minimum in the color distribution of stars with
$18 < V < 18.6$ defines the boundary between the VRC and the RGBB and
sets the red end of the color range. We define the blue end to
produce a symmetric color range about the peak of the red clump distribution.

To demonstrate that the VRC feature exists in the data for the different
filters, we plot in Figure 2
the distribution of the magnitude differences,
$\Delta m$, between the stars in this color range and the 
magnitude of the RC peak.
In all four bands, the $\Delta m$ distribution is
asymmetric,
and, ignoring the $U$ data for which the photometric uncertainties
are significantly larger, the distribution has a shoulder at 
$\Delta m \sim -0.9$. For comparison, we superpose the $I$-band $\Delta m$
distribution as a dotted line onto the $B$ and $V$-band distributions.
In contrast to the color-independence of the VRC,
the position of the RGBB relative to the RC changes 
among the different filters (see Figure 1).
Given that (1) the photometry from each of the
four filters is independent, (2) the amplitude of the photometric
errors is different in the four filters, (3) the extended
feature in the magnitude distribution is asymmetric to {\it bright}
magnitudes, (4) there is a minimum in the color distribution of stars
with 18 $< V < 18.6$ between the VRC and the RGBB, and (5) 
the VRC is fixed relative to the RC
in the $B$-, $V$-, and $I$-band
data while the RGBB moves,
we conclude that the VRC is a real feature and distinct 
from the RGBB. 

\noindent
{\it b) The VRC is due to clumpy extinction.}

Could a clumpy intervening dust distribution, either in the Galaxy or the
LMC, introduce sufficient scatter among
the magnitudes of the RC stars to produce the VRC?
The extinction required to produce the VRC (a $\sim$ 0.9 mag offset in $V$)
would produce E($B-V$) = 0.3 and E($B-I$) $\sim$ 1 mag (for
a standard extinction law; cf. Schild
1977). If such extinction were the cause of the VRC, then the 
magnitude offset, $\Delta m$, 
between the RC and the VRC would be 1 magnitude
greater in
$B$ than in $I$. From the comparison of the magnitude distributions
in Figure 2, it is evident that such large color differences are not
present. We conclude
that differential extinction is not responsible for the VRC.

\noindent
{\it c) The VRC is due to Galactic disk contamination.}

Could the VRC be due to contamination from nearby Galactic stars? 
Such a component would need to have the colors of the
LMC red clump stars, while being intrinsically significantly fainter, 
and to have some structure in the CMD in order to create a bump at the
position of the VRC. 
Red dwarf stars at the main sequence turnoff may be a viable candidate.
Main sequence K2V stars,
which have the $B-V$ color of the RC, would have to lie
at the unlikely distance of 2.3 kpc to match the 
magnitude of the VRC stars ($V \sim 18.3$). Even if 
there is such a population of stars, would
its distribution in the CMD have
a fairly distinct break to create the shoulder observed in
the $\Delta m$ distribution (Figure 2)? The main sequence turnoff,
if the turnoff is
located at $B-V \sim 0.9$ and at a $V$ magnitude that
is 0.9 mag brighter than the LMC RC, might create the required
break.
The oldest main sequence turnoff stars in the Galactic disk population
have an age of $\sim$ 10 Gyr 
(as suggested by the oldest white dwarf stars; Wood 1992).
Therefore, main sequence turnoff stars must be as red or bluer than 
10 Gyr old main sequence stars, which have $B-V = 0.62$ (Strai\u zys 1995). 
Because the RC has $B-V = 0.9$ and the VRC has
the same color as the RC to within 0.07 mag (see below),
main sequence turnoff stars cannot be responsible for the VRC.
We conclude that Galactic foreground stars are not responsible for the
VRC because (1) stars with the proper colors are not expected to have
any distinguishing distribution in the CMD and (2) the stars with
a possible distinguishing distribution have the wrong colors.

\noindent
{\it d) The VRC is due to stellar evolution.}

Could the VRC be a feature created by the evolution of red clump
stars, presumably upward from the RC to the VRC?
As RC stars evolve, they move within a limited, but noticeable,
region of the CMD (Seidel, Demarque, \& Weinberg 
1987; Sweigart 1987).
Models for the evolution of RC stars demonstrate that
their evolution proceeds primarily along the magnitude 
axis (\ie the color evolution can be relatively limited
for particular choices of stellar parameters).
Although these authors do not produce synthetic CMDs,
the evolutionary tracks they present indicate that the luminosity increase
necessary to populate the VRC (0.9 mag) is not caused by 
the stellar evolution of clump stars. The maximum change predicted 
by Sweigart for a metallicity, $Z$, of 0.01 (roughly appropriate for
the LMC; Westerlund 1997) is about 0.6 mag, with a typical value being 0.3 mag.
Synthetic $B-V,V$ CMDs
generated by Catelan \& de Freitas Pacheco (1996)
confirm this conclusion. 

Could the VRC stars be composed of 
a younger LMC stars that have a more luminous
red clump phase?
To examine whether another LMC stellar population can
create a second clump that is 0.9 mag brighter than the principal
red clump, but that has the same colors,
we examined Bertelli \etal 's (1994)
isochrones. We chose an isochrone with $Z$ = 0.008 and log(age) = 9.4 
as the best match to the giant branch morphology. Isochrones of
younger stars (log(age) = 8.6)
have a red clump phase that is $\sim$ 0.9 mag
brighter, but the difference in $B-I$ colors between the two red clumps
is about 0.2 mag. The observed $B-I$ color difference between the RC
and VRC is $\ltsim$ 0.07 mag (see below). 

Based on the failure of the stellar evolution models to produce
a population of stars that is brighter than the RC stars by 0.9
mag and that has the same $B-I$ color as the RC stars (within 0.07 mag), 
we conclude that neither stellar evolution within the
clump nor the presence of a younger stellar population is responsible for the
VRC. We caution that current stellar models, 
which include implementations of semiconvection, are complex and
not necessarily definitive, and that we explored a limited set of models.

\noindent
{\it e) The VRC is due to stellar blends.}

Could the blending of two stars into one apparent star create
VRC ``stars''?
In the Magellanic Cloud, stellar fields are crowded and some blends
undoubtedly occur. If the VRC ``stars'' are blends, then the
unblended components presumably have quite different photometric
properties.  The simplest way for a blend to have the same colors
as the RC is if the blended components are both RC stars. Such blends are
reasonable because the RC is the dominant stellar component at these 
magnitudes (see Zaritsky, Harris, \& Thompson 1997). Blends of RC 
stars with 
much fainter main sequence stars are irrelevant because they will not
alter the apparent luminosity of RC stars sufficiently. Blends with other
luminous stars (\eg young main sequence stars or red giants)
will produce ``stars'' with a wide range of colors. Therefore, we
consider only two 
types of RC-RC blends: (1) the random blend, in which
stars are physically unassociated but happen to lie sufficiently
close to the same line of sight, and (2) the correlated blend,
in which the two stars are physically associated.

We estimate the effect of 
random blended pairs of RC stars by adding artificial 
RC stars to our
images. The magnitude of the artificial stars is set equal to the centroid
of the RC distribution. We add 500 stars to a $V-$band 
subimage from our survey
that has roughly 17,000 identified stars using DAOPHOT's 
ADDSTAR procedure (Stetson 1987).
We then process the photometry for the image as was
originally
done for the true data 
(Zaritsky, Harris, \& Thompson 1997). The result from the
10 realizations of this test 
are shown in Figure 3, where we plot (1) the observed
distribution (solid line), (2) the 
resulting distribution of measured magnitudes for the simulated stars
(dotted line), and (3) 
the distribution of measured magnitudes if the simulated stars
are distributed along the line of sight in a disk with a vertical (i.e.
line-of-sight) scalelength of 5.3 kpc (dashed line), chosen to produce a close
match to the peak of the observed distribution.

The dashed line distribution is an overestimate of the effect
of blending for two reasons. First, we have adopted the
photometric errors from the original simulations. For the model
in which we distribute the RC stars along the line of sight, the 
relevant stars, in terms of the VRC, 
are those stars that are placed closer to us and are
therefore brighter than the RC centroid
(\ie at $\Delta m < 0$). These stars will be 
affected less by blending than in the original simulation because
they are now brighter.
Second, the density of simulated RC stars
is more than four times larger than in the real data, so RC-RC blends are
overrepresented. Even when overestimating the blending effect, 
we cannot reproduce the asymmetric distribution out to 
$\Delta m = -0.9$. We conclude that random blends are not responsible
for the VRC.

If the RC stars are clustered, then blends will be more common
than predicted in the random distribution model described above.
Unbound pairs of stars with velocity dispersions
$>$ 1 km sec$^{-1}$ will quickly ($\sim 2\times 10^5$ yrs) separate
by more than 1 arcsec and become unblended in the images.
Therefore, if tightly correlated RC blended pairs exist, they must
be bound binary systems.

There are empirical and theoretical arguments 
against the presence of a significant number
of RC binaries. We begin with an empirical argument.
In the upper panels of
Figure 4, we show the observed $\Delta m$ distribution for stars in $B$, 
$V$ and $I$. We 
subtract the RC contribution in two
different ways. In the first case, 
we model the RC distribution using a Gaussian
to represent the upper portion of the peak and an exponential to
represent the 
tail of positive $\Delta m$ values. 
We adjust the model parameters to provide the best fit both to
the upper portion of the distribution and to the positive tail. The
residuals
from this model are shown in the middle panels. In the second case, 
we assume that
the RC distribution is symmetric in magnitude
about the peak and use the faint
half of the distribution to define the RC contribution to the bright
half of the distribution. 
We do not apply this method to the $B$ data due to incompleteness
at $\Delta m_B \sim 1$. The residuals
from the symmetry model are shown in the bottom panels of Figure 4. 
Although neither method is ideal because of the assumption of a symmetric
RC distribution, both methods will most likely underestimate the
contribution of the RC at $\Delta m < 0$, because
contamination from some blends and younger stars probably extends
the bright tail of the RC distribution. 

We
measure the position of the peak of the residual distributions
using a parabolic fit within $\pm$0.3 mag of the peak. The
peak positions, as measured in all three filters, 
are within $-0.91$ mag $< \Delta m < -0.84$
mag. These measurements place the residual peak systematically at 
brighter magnitudes than 
the expected position of RC-RC binary
stars ($\Delta m = -0.75$ because the blended star is twice
as luminous). The
expected distribution of RC binaries 
is shown by the dashed lines (modeled to have the same shape
as the central RC distribution and height equal to the residual
distribution in the middle panels for comparison). These measurements also
demonstrate
that the relative colors of the RC and VRC do not differ by more
than 0.07 mag from $B$ to $I$.

There is a consistent offset between the position of the
observed $\Delta m$ distribution and that expected for a
RC-RC binary population, but systematic problems remain. The residual stellar
distribution is sensitive to the adopted shape of the RC star
distribution, which
is unknown. In addition, there is a general ``background"
distribution of stars throughout the CMD
that is not modeled. This background is unlikely to be critical
because any such background component will probably increase in density
toward 
fainter magnitudes, and its removal would only push the observed residual
peak to brighter magnitudes. Nevertheless, there are sufficient
systematic uncertainties that the 0.1 mag offset is only suggestive
evidence that these stars are not RC-RC binaries. 

A theoretical argument against large numbers of RC binaries
rests on the fine evolutionary timing required to get two stars
in the RC phase at the same time. 
The timing requirement can be converted into a statement about the
relative masses of the progenitor stars.
The bulk of RC stars have progenitors
with main sequence lifetimes in the 1 to 4 Gyr range (in order to
match the increase in the star formation efficiency in the last
few Gyr inferred
by several previous investigators for the LMC, cf. Gallagher \etal 1996)).
If two stars are going to reach the red clump phase at the same time,
then their main sequence lifetimes must differ by less than the red clump
phase lifetime.
Therefore, the maximum allowed percentage difference in their main 
sequence ages, $\sim 5\%$,  is given by 
the ratio of the red clump lifetime, $10^8$ years
(cf. Castellani, Chieffi, Pulone 1991),
to the main sequence lifetime, $t_{ms}$.
Using
the relationship that $t_{ms} \propto (M/M_\odot)^{-3}$
for stars somewhat more massive than the Sun (cf. Mihalas and Binney 1981),
the upper limit on the mass difference of the two progenitor stars is
2\%. Therefore, to account for the VRC with RC binaries implies
that roughly 10\% of all RC stars
are in binary systems in which the progenitor masses differed by
$\ltsim 2\%$. We reject this explanation due to the high degree
of fine tuning necessary in the binary star mass ratios.

\noindent
{\it f) The VRC is due to foreground RC stars}

Could the VRC stars be RC stars that are brighter than the
LMC RC because they are nearer to us? To investigate this hypothesis,
we have converted the magnitude differences for each star relative to the
LMC RC centroid into a distance by assuming that
all RC and VRC stars share the same absolute magnitude and that
the centroid of RC stars lies at the distance 
of the LMC (50 kpc; Feast \& Walker 1987). We have also
corrected the observed number counts at each distance
to a stellar density 
by accounting for the differential volume (our fixed area of sky
contains different volumes at different distances). The 
inferred line-of-sight stellar distributions are shown in the
top panels of Figure 5. The $B,V$ and $I$ distributions have a clear
shoulder at a distance of about 35 kpc.
Because the seeing is the worst in $U$,
the photometry is also the worst, and the features are broader and
weaker. Nevertheless, when we fit the simple model described
below, we find that the excess stellar density due to the VRC is
present in all four filters. The centroid of the VRC, as determined
from a parabola fitted to the stellar
density distribution between 30 and 40 kpc (and
between 25 and 45 kpc for the U band data due to the less pronounced
peak), is at 33.7, 35.0, 34.8, 34.1 kpc for $UBVI$, respectively.
The peak is at 33 kpc if one adopts $\Delta m = -0.9$ as the centroid
of the population and {\it does not}
correct the density for volume effects. The latter approach is appropriate
if the magnitude distribution around the VRC 
peak does not arise from differential distances relative to the VRC centroid.

The model shown in Figure 5 (solid line), 
is a combination of three components: (1) a Gaussian, with a
dispersion chosen to fit the peak of the distribution, (2) an
exponential, with a scalelength chosen
to generate the extended tail beyond 50 kpc, and (3) a power-law foreground component, with
an index chosen to
account for the sharp rise of objects at small distances. The
exponential is fit to the faint tail of the stellar distribution between 50
and 60 kpc. We ignore the data corresponding to inferred
distances greater than 60 kpc, because the
magnitude cuts originally applied to define the sample exclude such
distant RC stars.
The model exponential has the same
scale-length (4.5 kpc) for all four filters. This value should, for
the purposes here, be 
considered only as an empirical quantity that is a good fit
to the data, rather than necessarily
having physical significance. We adopt an exponent of $-$0.95 for 
the power-law component in all four filters, again as an empirical fit.
The color of the stars responsible for the power-law
component depends slightly
on magnitude, so the number 
represented at any distance in
Figure 5 is not a constant fraction of the total population. Therefore,
the exponent should not be interpreted physically.

{\it The agreement among distances measured in each of the four filters, and
among the amplitudes of the residual distributions in $B$, $V$, and $I$, is
consistent with 
the interpretation of the VRC as an intervening population in the 
line of sight to the LMC.}
We estimate the projected angular stellar density of such a 
foreground population relative
to that of the LMC by comparing the number of stars in the VRC with the number
of stars in the RC. This estimate is highly model dependent because both the 
RC and any foreground component 
must be subtracted from the observed distribution of stars. 
Nevertheless, we use the models just described
and estimate
that the VRC has a projected angular number density of $\sim$ 5 to 7\% of 
the LMC RC stars. This value is most likely an overestimate
for at least two reasons: (1) due to stellar evolution, blends,
and binaries (which all work to generate stars that are brighter 
than the RC centroid),
the RC is probably more extended toward $\Delta m < 0$
than the symmetric model implies 
and (2) the 
RGBB is likely to partially pollute the region of the VRC.

\subsection{A Corroborating Observation: The Lower Main Sequence}

If the VRC is a manifestation of a population between us and the LMC,
then the entire LMC CMD diagram should contain traces
of a parallel CMD, shifted in magnitude by about 0.9 mag. 
We have searched for
evidence of a parallel RGB, but are unable to draw any conclusions
due to the steepness of the RGB and our photometric uncertainties,
which would obscure a faint foreground population.
The upper main sequence is also not useful for this exercise,
because it is nearly vertical in our CMDs.
Our only remaining option is to examine
the lower main sequence, but the sensitivity limit and photometric
errors of our data preclude their use for such a study. Instead,
we examine HST data of the field population in the LMC
generously provided by J. Holtzman (the
data are discussed and analyzed by Holtzman \etal 1997). 

The $V-I,I$ CMD from Holtzman \etal\ is shown in 
Figure 6. There is a faint trace of a secondary main
sequence (it is visible over the range $0.6 < V-I < 1$ where the
main sequence stars are well below the turnoff magnitude and the photometric
errors are still small). Differential reddening
is not responsible for the apparent gap between the main
sequence and the secondary sequence, because the reddening 
vector is nearly parallel to the main sequence.
Models presented by Holtzman \etal (their
Figure 4) to study the star formation history of the LMC
illustrate that neither metallicity nor age 
variations between populations will populate the apparent secondary sequence.

There are two quantitative consistency checks on the hypothesis
that the secondary sequence is drawn from the same foreground
population as the VRC.
First, the secondary sequence should
appear $\sim$ 0.9 mag above the main sequence. Second, the secondary
population should have a projected angular number density of 
$\ltsim$ 5 to 7\% of that of the principal population. For
these tests, we define a fiducial main sequence by measuring the
mode of the stellar colors in magnitude bins for the main sequence
stars and by then fitting a second order 
polynomial as a function of color to the magnitudes. Our resulting
fiducial main sequence is $M_V = -3.26 + 19.66(V-I) -
9.99(V-I)^2$. For each star between $0.7 < V-I < 0.9$ and $ 3.5 < M_V <
7.25$, we evaluate the magnitude difference relative to the fiducial
main sequence. We set the blue color limit to avoid confusion with 
the subgiant branch at bright magnitudes 
and set the red color limit to avoid being seriously affected
by photometric errors.
The distribution of magnitude differences, 
for both the observations and simulated
data are in Figure 7 
(see Holtzman \etal for details regarding the simulation).

The observed data clearly contain a secondary peak 
at $\Delta m \sim -0.8$. The position of the residual
peak as predicted by the
VRC is shown by the arrow, with the differences in positions among the
four filters illustrated by the horizontal error bar. Any attempt
to subtract the contribution from the principal main sequence
will shift the secondary peak toward $\Delta m = -0.9$,
in even better agreement with the
position of the VRC (note that in Figure 2, before any RC
``background'' 
subtraction is attempted, the VRC is also closer to
$\Delta m = -0.8$).
No such peak is observed in the
simulations (the peak at $\Delta m = -$0.6 mag is not significant at
the 2$\sigma$ level, while the observed peak is significant at
$> 4 \sigma$). As for the VRC, determining the number of stars in
this secondary peak requires modeling of the primary peak.
After crudely modeling the contribution of the principal main
sequence as half of the signal seen in the secondary peak,
we estimate that the angular stellar density of the
secondary population is $\sim$4\% of the LMC's 
(if the contribution
from the LMC main sequence to the secondary peak
is negligible, then the percentage
increases to $\sim$ 8\%). {\it Therefore, both the position
and amplitude of the secondary peak seen in the main sequence population
are consistent with the values that
one would predict from the properties of the VRC. }

The observation of a brighter population of stars in another part
of the CMD greatly strengthens the case for an intervening population. 
and provides additional evidence against 
alternative
interpretations for the VRC. First, neither stellar evolution nor
metallicity effects can account for this parallel main sequence population 
(see Figure 4 of Holtzman \etal 1997). The presence of a younger
population (which might affect the VRC) would have no effect on this
portion of the main sequence. Second, blends are highly unlikely
to be responsible for the parallel main sequence. Unlike in the VRC
region of the CMD,
where the red clump stars dominate and red clump-red clump
blending will be the most common type of blend, there is no reason to expect
main sequence 
stars to blend with stars of exactly the same magnitude (as required to 
create a distinct peak at $\Delta m \sim$ 0.8 mag). 
Because the main sequence data come from HST observations and the 
VRC data come from  ground based observations,
it would be particularly unfortuitous for both data sets to have 
a similar blending problem for such different stellar
populations. Finally, the possibility of binary stars, which is
a much more likely explanation for the lower main sequence excess
population (because the timing constraints are not rigid), is not
a promising explanation of the VRC.
For example, Rubenstein
and Bailyn (1997) have observed the effects of binary stars on
the CMD of a core-collapse globular cluster, NGC 6752.
The signature they observe,
an asymmetric expansion of the main sequence toward redder and
brighter objects, is similar to that described here. However, if one invokes
this explanation for the lower main sequence excess population, 
another explanation is necessary for the
VRC. Such hybrid explanations for the two populations are certainly possible,
but become increasingly unappealing as more complications are added.

\section{Implications}

\subsection{For MACHO Microlensing Experiments}

The MACHO collaboration has measured an optical depth to
microlensing of $2.9^{+1.4}_{-0.9} \times 10^{-7}$ for
events between 2 and 200 days toward the LMC (Alcock \etal 1997). 
The subsequent interpretation of that result
relies on models of the stellar and dark matter halo distributions.
Any
irregularity in the density of lenses in the halo,
especially one that would be expected to correlate with the position
of 
the Large Magellanic Cloud, would seriously affect any interpretation.
Zhao (1997) demonstrated that a stellar stream from
a proto-LMC
that was originally
twice as massive as the current LMC
could provide the necessary microlensing optical depth.
Such models imply that MACHOS comprise at most a small
fraction of the Galactic halo.

We can straightforwardly estimate the optical depth to microlensing,
$\tau_{\mu}$,
of the putative intervening population toward the LMC using
\[\tau_{\mu} = 4\pi G \Sigma D_{LMC} x (1-x),\]
where $\Sigma$ is the surface mass density of the lenses, $D_{LMC}$ is
the distance to the LMC, which is taken to be 50 kpc (Feast \& Walker
1987),  and $x$ is
the ratio of the distances between the sources and the lenses.
For a population at 35 kpc, this equation can be rewritten as
\[\tau_{\mu} = 7.5 \times 10^{-9} \Sigma,\]
where $\Sigma$ is in units of $M_\odot/pc^2$. To estimate the
surface mass density of the foreground population, we assume that
its stellar population is identical to that of the LMC. 
Our observed
field spans a range in radius of $\sim$ 1 to 3 kpc (in the LMC disk plane
these radii correspond to 1.8 and 5.5 kpc for our adopted LMC
inclination of 33$^\circ$).
We estimate the average surface
mass density of the LMC using the rotation curve and assuming
that dark matter makes a negligible contribution to the
rotation curve at these radii.
Luks and Rohlfs (1992) present a rotation curve
for the disk of the LMC that has a roughly flat rotation curve
with a velocity of 70 km sec$^{-1}$ at these radii.
Adopting the simple spherical formula, $M = rv^2/G$, implies
a mass in this annulus of $4.0 \times 10^{9} M_\odot$. Projecting
this mass on the observed annulus from 1 to 3 kpc results in an
average projected 
surface mass density of 159 $M_\odot/pc^2$. If the foreground
population is 5\% of the 
projected angular surface number density of the LMC population (see 
\S2),
then the surface mass
density of the foreground component
would be
8 $M_\odot/pc^2$ if it was at the distance of the LMC. 
However, 1 pc$^2$ at 35 kpc corresponds to 2 pc$^2$
at 50 kpc, so the mass surface density of the inferred foreground population
is 16 $M_\odot/pc^2$. This surface mass density
translates to $\tau_{\mu} = 1.2 \times 10^{-7}$,
or about half of the measured value (but only 2$\sigma$ away using
only the uncertainty in the MACHO microlensing optical depth). 

We conclude that inferences regarding the halo MACHO population depend
critically on the characteristics of the putative foreground
population. Limits that constrain the foreground population
to well under 5\% of the LMC angular stellar density (in the region
of the LMC that we observed)
are required to reach the conclusion that the microlensing observations
necessarily imply the presence of purely halo MACHOS.
{\it Even if an alternate interpretation
for the observations discussed in this paper proves to be principally 
correct, we conclude that placing limits
on foreground populations within 20 kpc of the LMC is extremely complicated
and requires a detailed understanding of stellar evolution, binary
fractions, binary mass ratios, stellar correlation functions, differences
in the stellar populations among satellite galaxies, and
reddening.}

\subsection{For Dynamical Models of the Satellite Population}

The origin of the proposed foreground stellar population is an interesting
puzzle. There are at least three possibilities: (1) these
stars are debris from the interaction between the LMC and the 
Galaxy, (2) these stars are debris from the interaction between the
SMC and LMC, or (3) these stars belong to an unknown dwarf spheroidal
galaxy, which currently may or may not be internally gravitationally bound.
We briefly discuss possibilities (1) and (3), for which we have some 
additional constraints.

Is the observed population of intervening stars consistent with 
predictions of the position and density of the tidal tails produced
in models of the interaction between the LMC and the Galaxy?
First, we consider the position of the tidal streamer. Typical estimated
positions resulting from the tidal model for the origin of 
the gaseous Magellanic Stream place the inner stellar stream anywhere
from 30 to 45 kpc (Lin, Jones, \& Klemola 1995) ---
in agreement with our interpretation
of the observations (a density peak at $\sim 35$ kpc).
Second, we consider the density of stars along the hypothesized stream.
In \S2 we measured that the VRC has 
a projected angular surface density of 5 to 7\% that of 
the LMC. The HST data from observations of 
the lower main sequence support this value. The HST
measurement of the foreground stellar density
is important because it is obtained from
main sequence stars rather than RC stars, and so, it is less affected
by possible population differences between the foreground population
and the LMC stars.
Unfortunately, the surface
density is only a weak constraint on the models because the 
parameters of the progenitor satellite galaxy
are unknown. 

Assuming that the foreground population is a tidal streamer, 
we can estimate the mass of the 
original satellite galaxy
from the current surface density of the proposed foreground
component.
The stars escaping from a tidally
stripped galaxy follow an orbit similar to that of the parent galaxy 
(cf. Johnston, Hernquist, \& Spergel 1995).
If stripped stars from the LMC have
velocities that differ by 100 km sec$^{-1}$, 
these stars would wrap around the Galaxy at a radius $\sim$ 
35 kpc within 2.5 Gyr.
Along segments of these streams
of stellar debris, the local velocity dispersion perpendicular to the
``orbit"
is comparable to the virial velocity of the parent galaxy prior to the
tidal disruption (Oh, Lin, \& Aarseth 1995).
Consequently, the width
of the stream, normal to its orbital plane, is comparable to 
the size of the parent galaxy.  During the final stage of tidal
disruption, the total mass contained within the stream is comparable 
to that of the residual stars in the parent galaxy.  In this limit, 
the ratio of projected stellar surface density along the stream to 
that in the parent galaxy is equivalent to the ratio of the angular 
extent of the parent galaxy to 2$\pi$.  The disk of the current LMC 
extends over $\sim 14$ degrees (Westerlund 1997),
so a ring of material that has a total
mass equal to the current LMC would have an angular projected
surface density of 14/360, or 4\%.  Therefore, the
observed projected density of the foreground population is
comparable to that expected if the population is the tidal debris 
of an LMC-like satellite, or a proto-LMC with twice the current LMC mass.

One possible problem for a tidal stream model is the lack of
observed stars associated with the neutral hydrogen Magellanic Stream
((Br\"uck \& Hawkins 1983).
Some authors (Meurer, Bicknell, \& Gingold 1985; Moore \& Davis 1994) 
have proposed models in which the gas is removed from the LMC by
ram pressure stripping, in which case stars are not expected to be in
the Stream. However, previous searches have focused on bright, early-type
stars which will not be present if the gas
was pulled out over 1 Gyr ago and there was no subsequent star formation.
A few carbon stars
with appropriate magnitudes have been
found in the Stream (Demers, Irwin, \& Kunkel 1993).  However, the number of
carbon stars detected is too small to make accurate estimates of the Stream's
stellar content.  Because the intrinsic nature
(and hence luminosity) of these stars is ambiguous, their distance cannot be
reliably determined. Even if low mass stars are not found in the Stream,
one might be able to account for the displacement of stars and gas
by appealing to a low density gaseous Galactic halo that generates a mild drag 
on the neutral hydrogen.

Now we consider the possibility that the proposed
foreground population is associated with a dwarf
galaxy along the line of sight.
We assume that the foreground stars, including the VRC, have a similar
stellar luminosity function and mass-to-light ratio as those in
the LMC. Therefore, 
the stellar surface mass density of the foreground population is $\sim$ 16
$M_\odot/pc^2$ (see above).
The surface 
number density of the VRC is not observed to vary over our field
(2$^\circ$, or 1.2 kpc at 35 kpc).
If a similar spatial extent is assumed along the line of sight,
the inferred mass density is 0.01 $M_\odot/ pc^{3}$.
This density is a factor of five smaller than the {\it minimum}
value of the central dark matter density found for the Draco and Ursa Minor
dwarf spheroidals (models with isotropic velocity distributions
predict mass densities of $\sim$ 1 $M_\odot /pc$; Pryor
\& Kormendy 1990). Therefore, if the mass-to-light ratio in
the foreground population is similar to that in Draco and 
Ursa Minor ($\sim 70$; Olszewski, Aaronson, \& Hill 1995), then
this population has a mass density comparable to the centers of 
dwarf spheroidal galaxies.
At their present Galactic distance ($\sim 69$ and 86 kpc,
respectively; 
Cudworth \etal 1986; Nemec \etal 1988; Nemec 1985), 
Draco and Ursa Minor
would be tidally disrupted by the Galaxy if they had no dark matter
(Faber \& Lin 1983, Oh, Lin, \& 
Aarseth 1993). Therefore, the foreground population, which is at 
half the Galactocentric distance,
would also be tidally disrupted unless it has substantial
dark matter. If the hypothesized foreground population is traced over a larger
radial range, then the inferred mass density would decrease (since
the object would be thicker as well as wider) and even more dark matter
would be required to keep it gravitationally bound.

\section{Resolving the Question}

There are several potential
ways to 
discriminate between the various hypotheses for the origin of the
VRC and the possibly related main sequence population.
Because certain
explanations (\eg blends, binaries, younger stellar populations)
predict direct
correlations between either the number of RC stars or
OB stars and the number of VRC stars,
observations of fields farther from the center of the LMC could verify or
refute these hypotheses. 
We have attempted this test within our survey area by counting 
both RC and VRC stars on a 10 by 7 rectangular grid,
statistically subtracting
the RC component, and examining the scatter 
plot between RC and VRC number densities. Although
we observe a slight anticorrelation between the two, supporting
the hypothesis that the RC and VRC are unassociated
and that the VRC is in the LMC foreground, the uncertainties
in the statistical correction are sufficiently large to mask a
correlation if present.
If the VRC stars do not correlate
with the number of RC stars, then a comparison of the number of VRC
stars in outer LMC fields along the LMC orbit, to the number in fields 
perpendicular to the orbit, 
would establish whether the VRC is spatially distributed 
in a tidal stellar stream.

Other tests involve looking for variable stars in the microlensing
databases. 
For example, the MACHO collaboration excludes the possibility 
of a dwarf galaxy along the line of sight
between us and 30 kpc based on the analysis of $\sim$6000
RR Lyrae stars (Minniti priv. comm).  These investigators avoid the region
within 20 kpc from the LMC due to the ambiguities in
disentangling reddening, intrinsic scatter, and distance effects.
Even without these problems, it is difficult to predict the 
magnitude of the expected signal in the RR Lyraes from the foreground
population suggested by the VRC.
There are
several unknown factors that come into play. First, the
ratio of the foreground to LMC populations can vary due to clumpiness
in the foreground component or
to variations in the LMC surface density.
For example, the microlensing
surveys have been more concentrated toward the LMC bar than our
observations, so the foreground population in these data would be a smaller
fraction of the total number of stars. Second, old stellar populations,
such as RR Lyrae,
generally have fairly steep radial density gradients ($\rho \propto
r^{-3}$ for Galactic RR Lyrae (Oort \& Plaut 1975) and 
$r^{-3.5}$ for globular clusters (Harris
1976)) so that
tidally stripped material may have a lower RR Lyrae/main sequence
ratio than the central LMC fields. Third, if
the foreground population is not
related to the LMC, then its RR Lyrae fraction is an unknown quantity
and may be lower, or higher, than that in the LMC. 

Radial velocity observations
might be more successful at discriminating between models for the VRC.
The radial velocity distribution
of the proposed foreground stars might be expected to be significantly different
than those of LMC RC stars.  If VRC stars are members of a dwarf galaxy
between the LMC and the Galaxy, their mean velocity may differ from
that of the LMC RCs by up to 200 km s$^{-1}$ with a relatively small
dispersion.  If the VRC stars are tidal debris from the LMC, they would have 
a relatively small mean velocity with a dispersion of up to 200 km
s$^{-1}$. If the magnitude of VRC stars 
is a precise distance indicator, it may 
correlate with the radial velocity of the VRC star. Differences between
the velocity distribution of the VRC and RC stars 
would support the idea that the VRC stars
are not in the LMC. A velocity distribution consistent with that of  
the LMC would be inconclusive.

\section{Summary}

We identify a previously unknown extension of red clump (RC) stars toward
brighter magnitudes that lies $\sim$ 0.9 mag above the red clump 
centroid in the
CMD of the Large Magellanic Cloud. 
The extension
appears in the photometry of all four filters, thereby confirming it
as a real feature in the CMD. The color of the extension is 
the same as the RC to within $\sim$ 0.07 mag over $B$, $V$, and $I$.

We explore various explanations of this feature, including
unresolved stellar blends (both random and correlated), stellar evolution,
composite stellar populations, 
patchy extinction, a Galactic disk population, and an intervening
stellar population at a distance of about 35 kpc.
The latter explanation is the most consistent with the
data. 

We found corroborating evidence for a foreground population in the
HST CMD of field stars in the LMC provided by Holtzman (Holtzman \etal
1997). These data show a distinct sequence of stars that is parallel to,
but brighter than, the bulk of the main sequence stars by $\sim$ 0.8 mag.
The magnitude
offset and the relative projected density with respect to the principal
LMC population are in agreement with the values measured from the vertically extended
red clump (VRC) stars . Although there are explanations other than a
foreground population for 
the VRC and the parallel lower main sequence, a single alternative 
explanation cannot reproduce both observations. Therefore, the
similarity in the properties of the VRC and parallel main sequence suggests
the existence of a foreground population at $\sim 35$ kpc.

If the intervening population exists, then
what are the
implications for (1) the interpretation of the microlensing
rates toward the LMC and (2) the origin
of this population? We estimate the microlensing optical depth
of this population to be $\sim 1.2 \times 10^{-7}$. This optical
depth is 
about one-half of the observed value (Alcock \etal 1996), 
but only 2$\sigma$ from the observed value. 
Given the large uncertainties in the 
measured optical depth and in our measurements of the foreground
surface mass density,
it is plausible that the entire microlensing rate can be accounted
for by normal stellar populations
and that there is no need to invoke purely halo MACHOs.

We speculate that this population originates 
from the LMC-Galaxy interaction, the LMC-SMC interaction, or
from an intervening dwarf galaxy (that is possibly
tidally disrupted).
Numerical simulations of the LMC-Galaxy interaction
(Lin, Jones, \& Klemola 1995) predict that a tidal tail
might lie between 30 and 45 kpc (consistent
with the distance of 35 kpc inferred from the peak of the VRC). 
The surface density constraints
from such models
are weak, because they depend on the unknown mass of the progenitor.
However,
if the original LMC had about twice its current stellar mass,
the projected surface density predicted by a simple calculation
is in agreement with the observations.

The existence of the VRC, and of its sister population near
the lower main sequence, is clear, but the interpretation of these
components is still ambiguous. The simplest explanation is that
a stellar population lies in front of the LMC. This population may be 
the signature of a tidal tail from the LMC and may
be responsible for the LMC microlensing events. In any case, we conclude
that it is exceedingly 
difficult from the CMD alone to rule out a foreground
population of stars at the few percent level. As a result, the issue of
possible intervening populations along the line-of-sight and within
$\sim$ 20 kpc of the Large Magellanic Cloud is entirely open. Therefore,
the observed microlensing rate toward the LMC should not yet
be interpreted as
evidence for a purely halo MACHO population, and
second order results, such as the mass distribution of halo MACHOS,
should be viewed with extreme caution.
Fortunately, several
tests of this hypothesis are feasible and should allow some progress
on this issue within the next few years.

\vskip 0.5in
\noindent
Acknowledgments: DZ gratefully acknowledges financial support from
a NASA LTSA grant (NAG-5-3501) and an NSF grant (AST-9619576), 
support for the construction of the GCC
from the Dudley Observatory through a Fullam award and a seed grant
from the Univ. of California for support during
the inception of MC survey.
DZ also thanks the Carnegie Institution
for providing telescope access, shop time, and other 
support for the MC survey. Finally, we thank A. Gould, P. Guhathakurta,
and D. Minniti for helpful discussions and A. Zabludoff for a careful
reading of a preliminary draft.

\clearpage
\centerline{References}
\refbook{Alcock, C. \etal 1997 preprint (astro-ph/9606165)}
\aasup{Bertelli, G., Bressan, A., Chiosi, C., Fagotto, F., \&
Nasi, E. 1994}{106}{275}
\aa{Br\"uck, M.T. \& Hawkins, M.R.S. 1983}{124}{216}
\apjsup{Castellani, V., Chieffi, A., \& Pulone, L. 1991}{76}{911}
\pasp{Catelan, M., \& De Freitas Pacheco, J.A. 1996}{108}{166}
\aj{Cudworth, K.M, Olszewski, E.W. \& Schommer, R.A. 1986}{92}{766}
\mn{Demers, S., Irwin, M.J., \& Kunkel, W.E. 1993}{260}{103}
\apjlett{Faber, S.M. \& Lin, D.N.C. 1983}{266}{17}
\annrev{Feast, M., \& Walker, A.R. 1987}{25}{345}
\aa{Fusi Pecci, F., Ferraro, F.R., Crocker, D.A., Rood, R.T., \&
Buonanno, R. 1990}{238}{95}
\apj{Gallagher, J.S., \etal 1996}{466}{732}
\mn{Hatzidimitriou, D. \& Hawkins, M.R.S. 1989}{241}{667}
\aj{Harris, W.E. 1976}{81}{1095}
\aj{Holtzman, J. \etal 1997}{113}{656}
\mn{Ibata, R.A., Gilmore, G., \& Irwin, M.J.}{277}{781}
\apj{Johnston, K.V., Hernquist, L., \& Bolte, M. 1996}{465}{278}
\apj{King, C.R., Da Costa, G.S., \& Demarque, P. 1985}{299}{674}
\apj{Kunkel, W.E. 1979}{228}{718}
\apj{Lin, D.N.C., Jones, B.F., \& Klemola, A.R. 1995}{439}{652}
\aa{Luks, Th., \& Rohlfs, K. 1992}{263}{42}
\refbook{Lynden-Bell, D. 1982, Observatory, 102, 202}
\apjlett{Majewski, S. 1994}{431}{17}
\apjlett{Mateo, M. \etal 1996}{458}{13}
\apjlett{Mathewson, D.S., Schwarz, M.P., Murray, J.D., 1977}{217}{L5}
\refbook{Meurer, G.R., Bicknell, G.V., \& Gingold, R.A. 
1985, {\it PASAu}, 6, 2}
\refbook{Mihalas, D., \& Binney, J. 1981, {\it Galactic Astronomy},
(W.H. Freeman \& Co.: San Francisco)}
\mn{Moore, B., \& Davis, M. 1994}{270}{209}
\aj{Nemec, J.M. 1985}{90}{204}
\aj{Nemec, J.M., Wehlau, A., \& del Olivares, C.M. 1988}{96}{528}
\apj{Oh, K.S., Lin, D.N.C., \& Aarseth, S.J. 1995}{442}{1420}
\aj{Olszewski, E.O., Aaronson, M., \& Hill, J.M. 1995}{110}{5}
\aa{Oort, J.H., \& Plaut, L. 1975}{41}{71}
\aj{Pryor, C., \& Kormendy, J. 1990}{100}{127}
\apj{Rubenstein, E.P. \& Bailyn, C. D. 1997}{474}{701}
\pasp{Sahu, K. 1994}{106}{942}
\aj{Schild, R.E. 1977}{82}{337}
\apjsup{Seidel, E., Demarque, P., \& Weinberg, D. 1987}{63}{917}
\apj{Stanek, K.Z. \etal 1997}{477}{163}
\pasp{Stetson, P.B. 1987}{99}{191}
\refbook{Strai\u zys, V. 1995, {\it Multicolor Stellar Photometry},
(Pachart Publishing House: Tucson)}
\refbook{Sutherland, W., et al. 1997, in ``Identification of Dark 
Matter'', in press.}
\apjsup{Sweigart, A. 1987}{65}{95}
\refbook{Westerlund, B.E. 1997, {\it The Magellanic Clouds},
(Cambridge Univ. Press: Cambridge)}
\apj{Wood, M.A. 1992}{386}{539}
\pasp{Zaritsky, D., Shectman, S.A., \& Bredthauer, G. 1994}{108}{104}
\refbook{Zaritsky, D., Harris, J., \& Thompson, I. 1997, in press}
\refbook{Zhao, H. 1997 preprint (astro-ph/9703097)}

\vfill\eject
\centerline{\bf Figure Captions}

\medskip
\noindent
\figcaption{}{Hess diagrams of the region including the red clump
and giant branches in $B$, $V$, and $I$ vs our modified color
measure (\cf \S2). The images are generated by creating a
two-dimensional Gaussian with width and height defined by
the observational uncertainties for each star, and then summing
all of those Gaussians. In the upper panel (for the $B$-band
data), we have placed vertical bars to indicate the color
selection used to isolate the VRC and RC stars from the
other populations. In the middle panel (for the $V$-band),
we have labeled the various components discussed in the text.
The vertical axis in each panel spans 3.5 mag. The diagram contains
about 70000 stars.\label{Figure 1}}

\medskip
\noindent
\figcaption{}{The distribution of magnitudes relative to the magnitude
of the RC centroid. For each of the four filters we show the
distribution
of stellar magnitudes within the color bin $2.85 < C < 3.57$. In
the $B$ and $V$ panels, we have overplotted the $I$-band distribution
for comparison as the dotted line. The
$B$, $V$, and $I$ panels also have a vertical line showing 
the position of the VRC centroid 
($\Delta m \sim -0.9$).\label{Figure 2}}

\medskip
\noindent
\figcaption{}{The effect of random blends. We summarize the results of
our artificial star tests. The solid smooth curve the $V$-band
distribution is from Figure 2. The dotted line shows the result of 
adding 5000 stars with the magnitude of the red clump centroid to
the survey images and remeasuring the magnitudes (\ie for no errors,
the distribution would be a delta function at $\Delta m = 0$).
The dashed-line histogram shows the results of expanding the simulated
distribution by modeling the RC stars as having an exponential
distribution along the line-of-sight with a scaleheight of 5.3 kpc
(selected to provide a good empirical match to the observed
distribution). For two reasons, this test is expected to be an
overestimate of the effect of blends (see text).\label{Figure 3}}

\medskip
\noindent
\figcaption{}{The $\Delta m$ distribution for the
$B$, $V$, and $I$ photometry. In the upper panels, we present the
data (solid line) and the Gaussian+exponential models as described
in the text (dashed line). In the middle panels, we present the
residuals (solid line) from the models drawn in the upper panels.
Also in the middle panels, we present for comparison the expected distribution
of RC-RC blends, normalized to the height of the observed residuals
and to the width of the principal RC peak. In the lower panels, we
present the residuals (solid line) for a second RC model, the symmetry
model (see text). Again, the expected distribution of RC-RC blends
is plotted for comparison (dashed line).\label{Figure 4}}

\medskip
\noindent
\figcaption{}{The line of sight density distribution of stars. The magnitude
distribution
of stars  within
the color bin described in \S2 is converted to a distance distribution
assuming that all stars have the same absolute magnitude and that 
it is equal to that of the centroid of 
the RC distribution. The density at each radius is corrected for the
volume surveyed at that distance. The upper panels show the
distribution
for photometry in each filter (dotted lines) and the model
(see text; solid line). In the lower panel, we show the
residual after subtracting the model. The dotted line is the $U$
data, short-dash/long-dash line is the $B$ data, short-dash line
is the $V$ data, and long-dash line is the $I$ data.\label{Figure 5}}

\medskip
\noindent
\figcaption{}{The $V,I$ color-magnitude diagram of the LMC field 
population from the data of Holtzman \etal (1997). The secondary
sequence is most clearly visible at about $V-I = 0.8$ and $V = 5.3$. 
\label{Figure 6}}

\medskip
\noindent
\figcaption{}{The line of sight density distribution of main sequence
stars from Figure 6, and the Holtzman \etal (1997) simulation of 
this population calculated as described in the text. The arrow
indicates
the position of the foreground population as expected from the 
VRC centroid. The horizontal errorbar attached to the arrow
indicates the spread in positions derived for the VRC from the 
data in different filters. \label{Figure 7}}

\setcounter{figure}{0}
\begin{figure}
\caption{}
\plotone{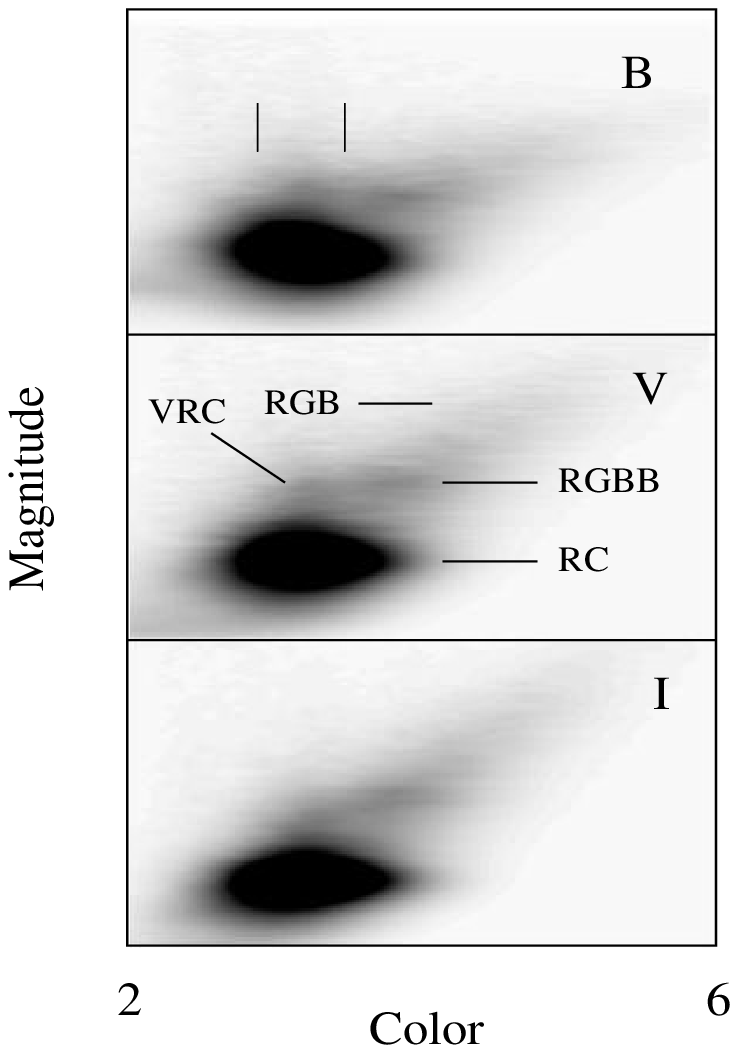}
\end{figure}

\begin{figure}
\caption{}
\plotone{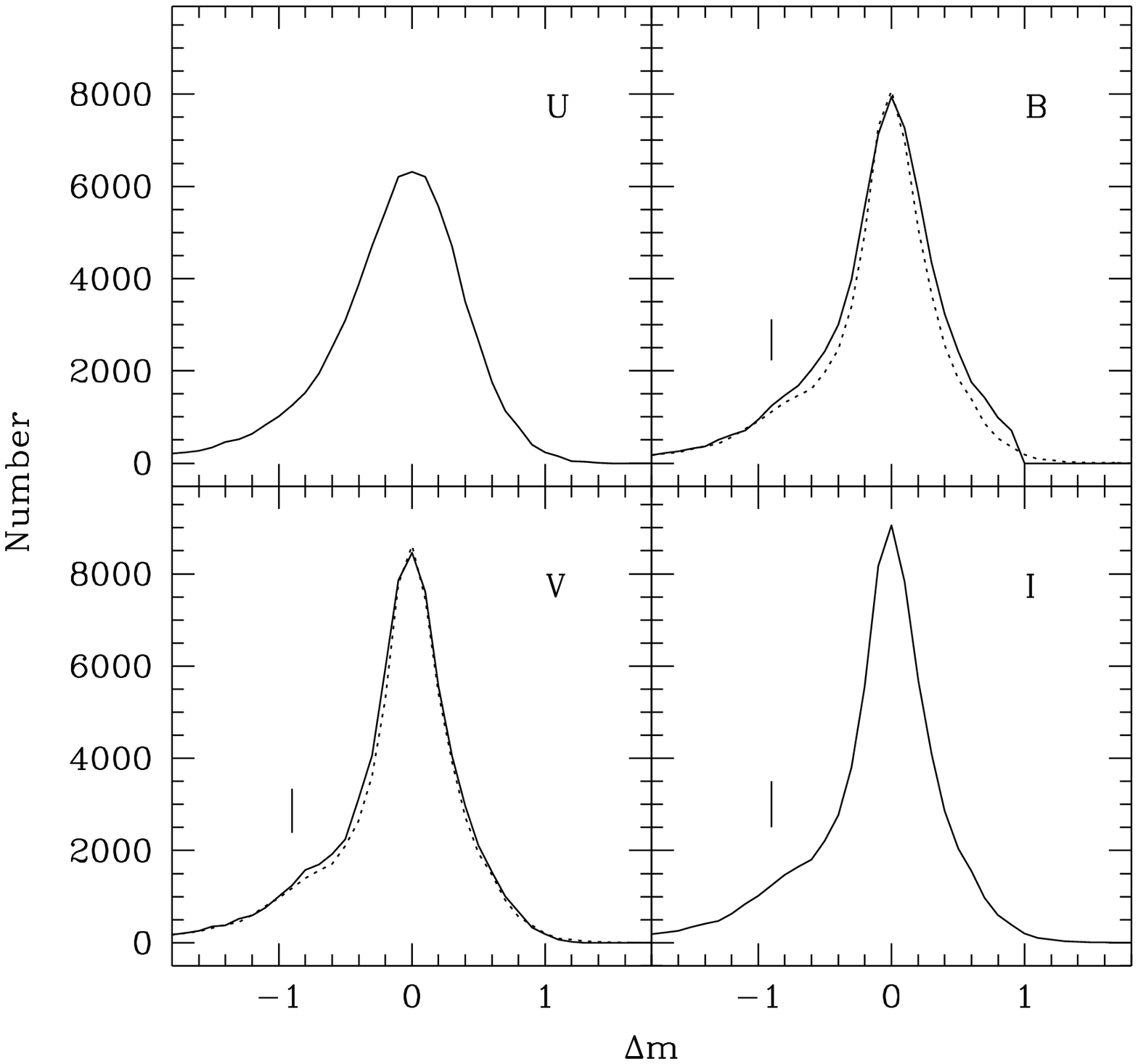}
\end{figure}

\begin{figure}
\caption{}
\plotone{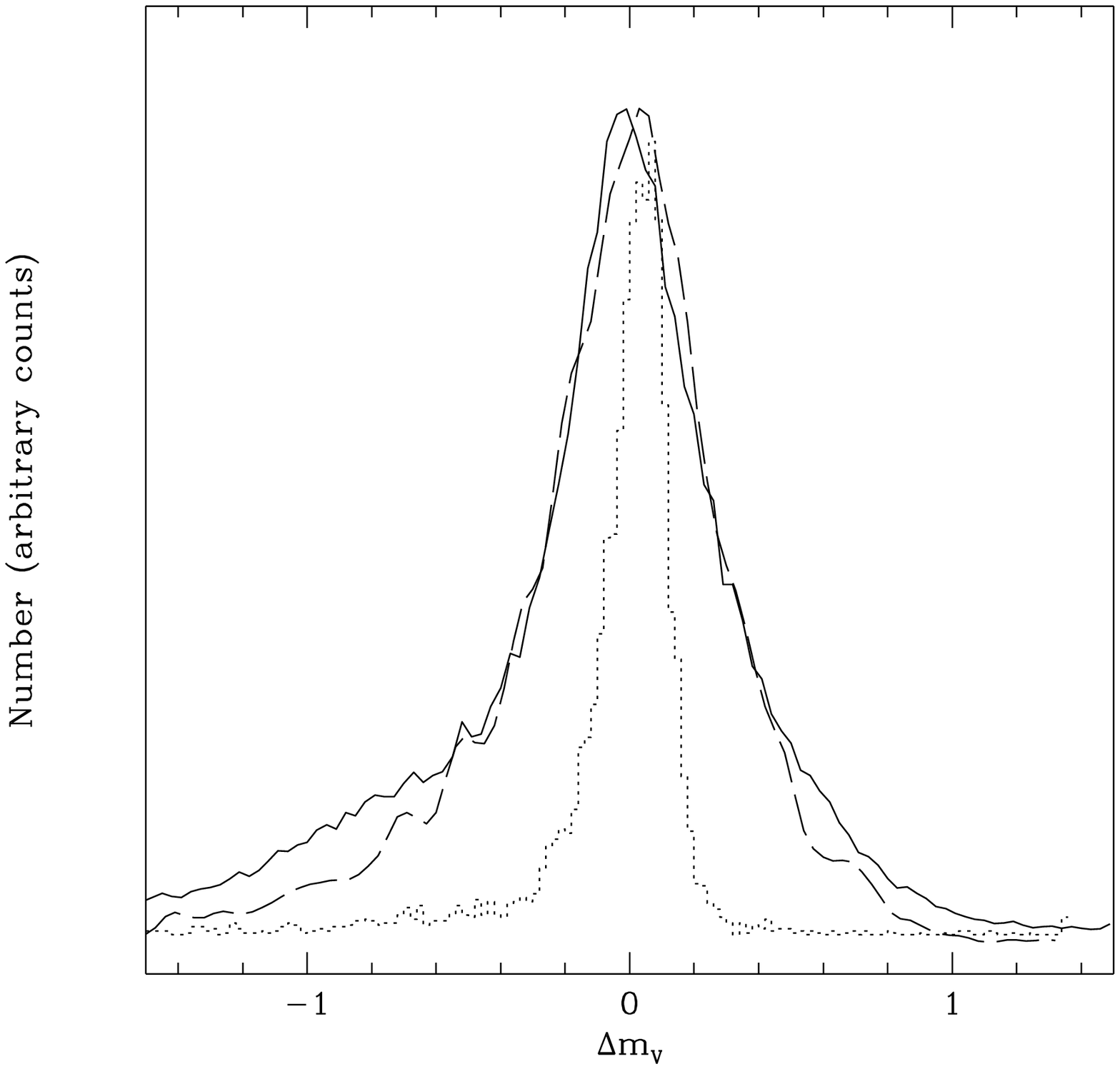}
\end{figure}

\begin{figure}
\caption{}
\plotone{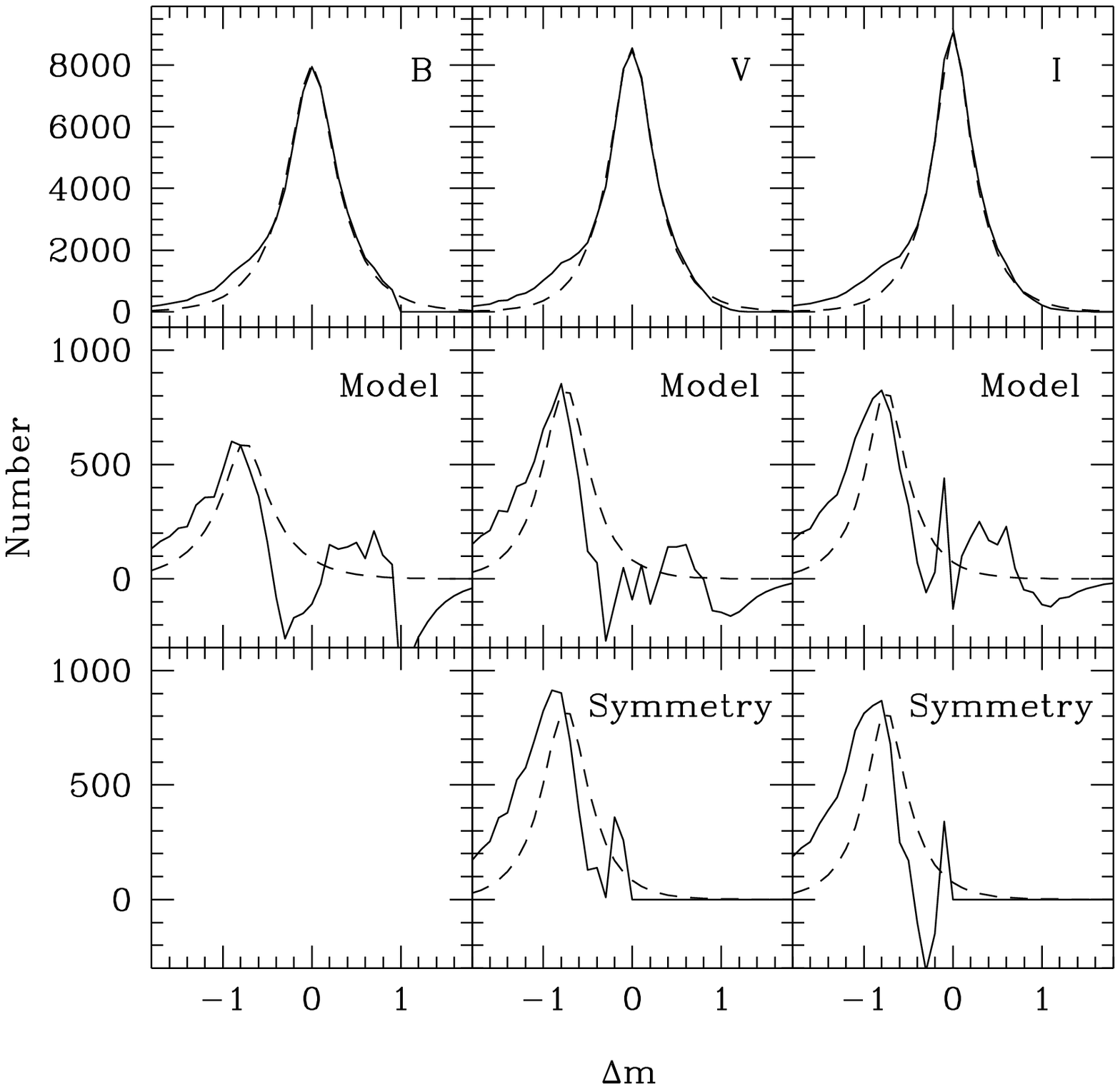}
\end{figure}

\begin{figure}
\caption{}
\plotone{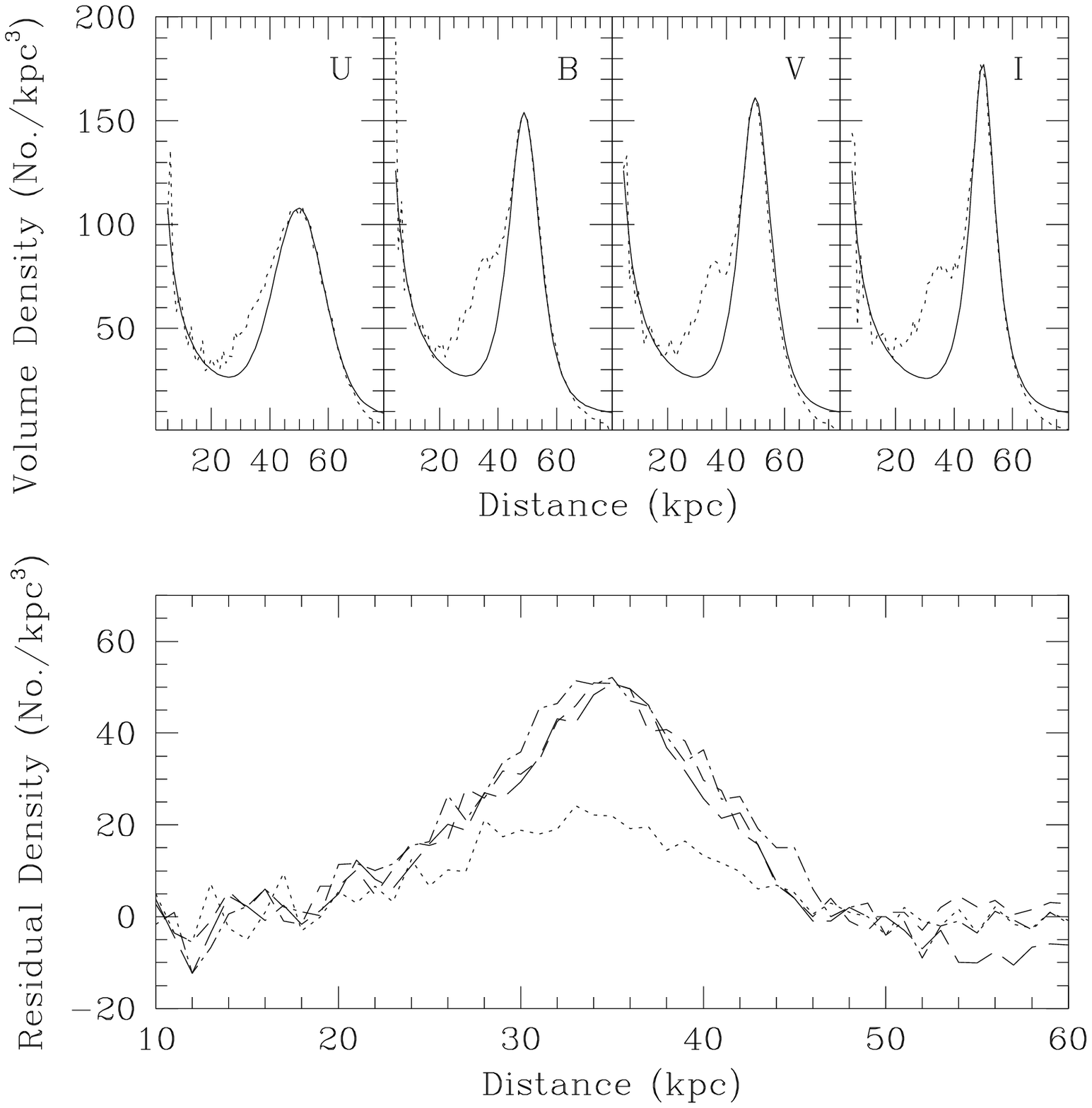}
\end{figure}

\begin{figure}
\caption{}
\plotone{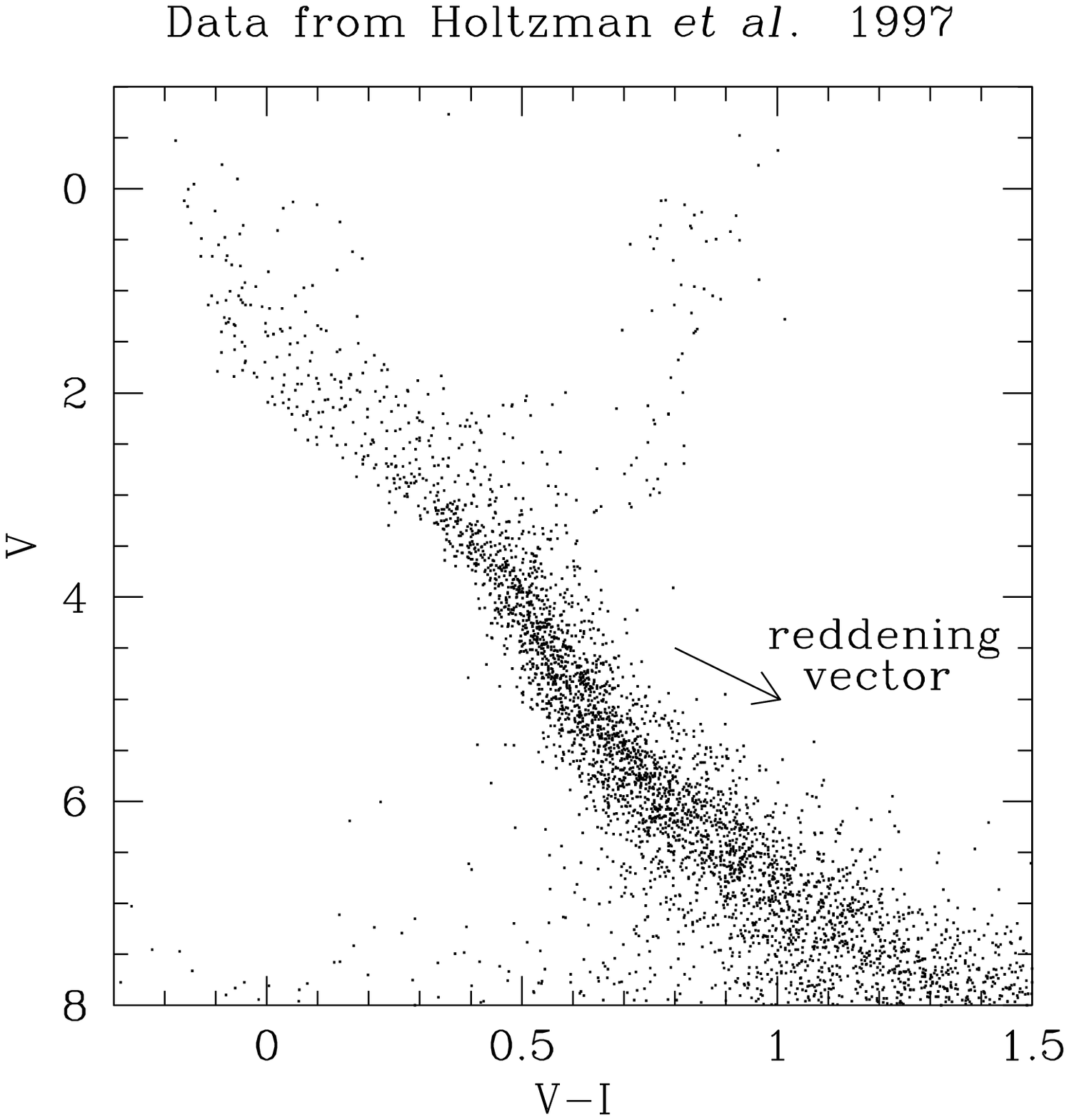}
\end{figure}

\begin{figure}
\caption{}
\plotone{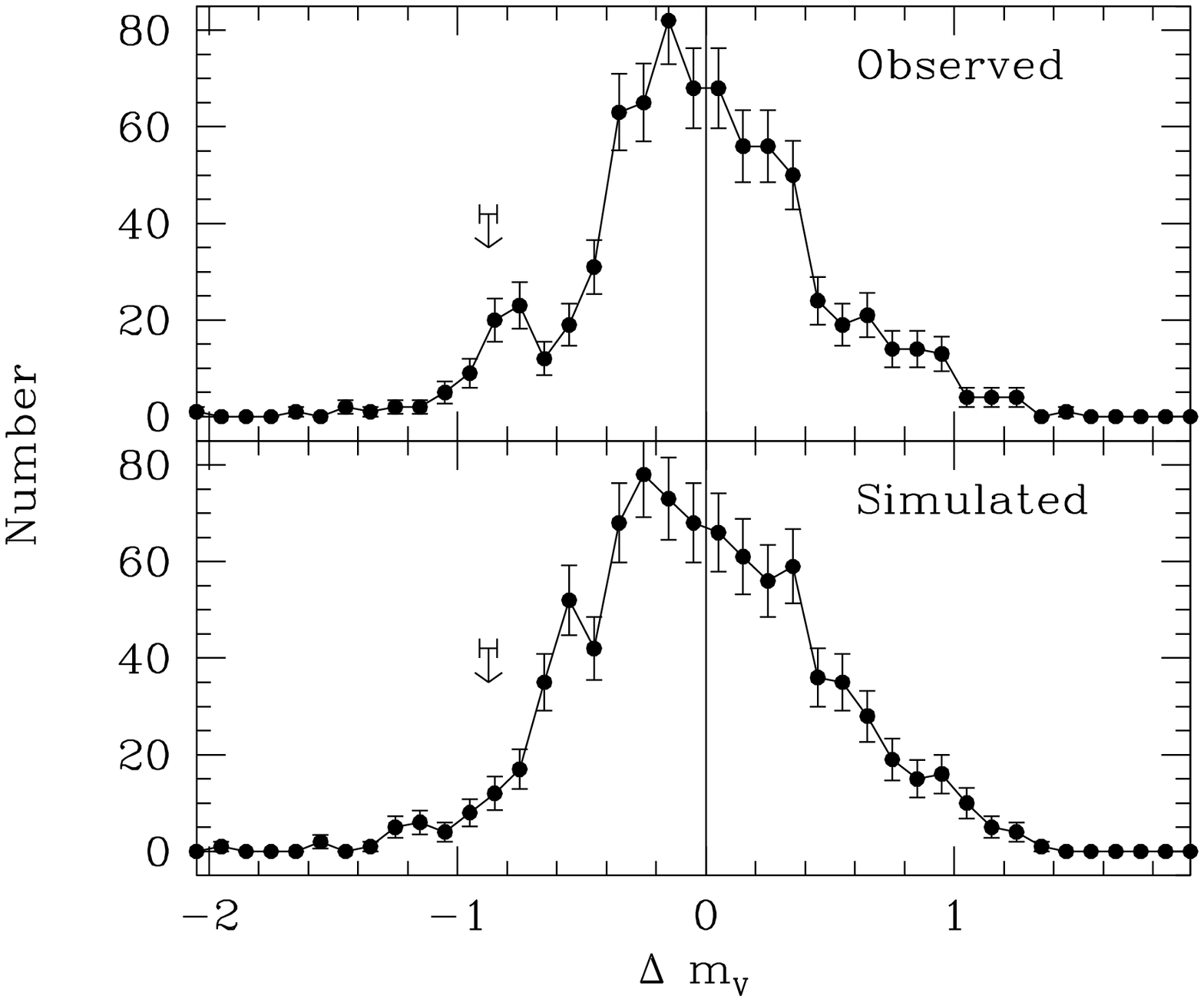}
\end{figure}

\end{document}